\documentclass[11pt]{article}
\usepackage{amsmath}
\usepackage{amsfonts}
\usepackage[dvips]{epsfig}

 \advance \textwidth by -2in
 \advance \textheight by -3in
\def\##1{{\bf #1}}
\def\=#1{\underline{\underline{#1}}}

\def\+#1{\underline{\bf #1}}
\def\*#1{\underline{\underline{\bf #1}}}

\def\eps{\epsilon}

\def\.{\mbox{ \tiny{$^\bullet$} }}

\def\le{\left(}
\def\ri{\right)}
\def\les{\left[}
\def\ris{\right]}
\def\lec{\left\{}
\def\ric{\right\}}

\def\c#1{\cite{#1}}
\def\l#1{\label{#1}}
\def\r#1{(\ref{#1})}

\begin{document}
\vskip 0.4cm

\noindent
{\bf PLANE WAVES WITH NEGATIVE PHASE VELOCITY IN ISOTROPIC\\ CHIRAL MEDIUMS
  }
\vskip 0.2cm

\noindent  {\bf Tom G. Mackay}
\vskip 0.2cm

\noindent {\sf School of Mathematics, University of Edinburgh\\
\noindent James Clerk Maxwell Building, King's Buildings, Edinburgh EH9 3JZ, United Kingdom}
\vskip 0.4cm

\noindent {\bf ABSTRACT:}
The propagation of plane waves  in an isotropic chiral medium (ICM)
is investigated.
Simple conditions are derived~---~in terms of the constitutive parameters of the
ICM~---~for the phase velocity to be directed opposite to the direction of power flow. It is demonstrated
that phase velocity and power flow may be  oppositely directed provided that the magnetoelectric
coupling is sufficiently strong.


\vskip 0.4cm

\vspace{10mm}

\noindent{\bf 1. INTRODUCTION}

Homogeneous mediums in which plane waves propagate  with phase
velocity directed opposite to the direction of power flow are
known as negative phase velocity (NPV) mediums. The archetypal NPV
medium is the lossless, isotropic, dielectric--magnetic medium,
characterized by the scalar permittivity $\eps $ and scalar
permeability $\mu $, with $\eps < 0$ and $\mu < 0$ being satisfied simultaneously.
In the 1960's Veselago predicted that such
a medium would exhibit
 a range of interesting and potentially useful
electromagnetic phenomenons, such as negative refraction, inverse
Doppler shift and inverse \u{C}erenkov  radiation \c{Veselago68}.
Recent experimental observations involving the microwave
illumination of certain composite metamaterials are supportive
Veselago's predictions \c{SSS}--\c{NPV_expt3} and have sparked an
intensification of interest in this area \c{LM_MOTL}.

In reality, isotropic dielectric--magnetic mediums are
characterized by complex--valued constitutive parameters $\eps$
and $\mu$. Phase velocity and power flow are oppositely directed
in such a medium provided that the inequality \c{LMW2}
\begin{equation}
\frac{\mbox{Re} \, \lec \eps \ric}{\mbox{Im} \, \lec \eps \ric} +
\frac{\mbox{Re} \, \lec \mu \ric}{\mbox{Im} \, \lec \mu \ric}  <
0\, \l{NPV_em}
\end{equation}
is satisfied at a given frequency,
where $\mbox{Re} \, \lec \. \ric$ and $\mbox{Im} \, \lec \. \ric$
denote the real and imaginary parts, respectively. Greater scope
for NPV propagation is provided by more complex mediums, as has
been demonstrated elsewhere for anisotropic \c{HC02,Kark} and
bianisotropic \c{ML_PRE} mediums.
 The present
communication concerns the generalization of the NPV condition
\r{NPV_em},
 within the regime of isotropic mediums. Specifically, we
consider the propagation of plane waves in
 the most general linear isotropic medium, namely the
isotropic chiral medium (ICM).
It is shown that
the introduction of magnetoelectric coupling extends the possibilities for NPV planewave
propagation in isotropic mediums.

\noindent{\bf 2. ANALYSIS}

Let us
consider the propagation of planewaves with field phasors\footnote{Vectors are displayed in boldface, with
  the symbol $\hat{}$ denoting a unit vector.}
\begin{equation}
\left.\begin{array}{l}
\#E(\#r) = \#E_{0}\, \exp (i k \, \hat{\#k} \. \#r )\\[5pt]
\#H(\#r) = \#H_{0}\, \exp (i k \, \hat{\#k} \. \#r )
\end{array}\right\},
\l{pw}
\end{equation}
where $| \hat{\# k}| = 1$,
in the isotropic
 chiral medium (ICM) described by  the Tellegen constitutive relations \c{Beltrami}
\begin{equation}
\left.
\begin{array}{l}
 \#D (\#r) = \eps\#E (\#r) + i \xi \#H (\#r) \,\\ [5pt]
 \#B (\#r)= - i \xi \#E (\#r) + \mu \#H (\#r) \, \l{cr}
\end{array}
\right\}.
\end{equation}
The scalar constitutive parameters $\eps$, $\xi$ and $\mu$ are complex--valued in general, but
for a lossless ICM we have $\eps, \mu, \xi \in \mathbb{R}$.

The wavenumbers $ k = k_R + i k_I $, with $k_R, k_I \in \mathbb{R}$,
are calculated from  the
  planewave dispersion relation
\begin{eqnarray}
\mbox{det} \, \les \, \=L(i  k \, \hat{\#k}) \, \ris  = 0 , \l{disp}
\end{eqnarray}
which arises from
 the  vector Helmholtz equation
\begin{eqnarray}
\=L(\nabla) \. \#E_0 = \#0 , \l{Helm}
\end{eqnarray}
wherein $\=L$ is the 3$\times$3 dyadic differential operator
defined as
\begin{eqnarray}
\=L(\nabla) = \nabla \times \nabla \times \=I - 2  \omega \xi
\nabla \times \=I
- \omega^2 \le \eps \mu - \xi^2 \ri \,. \l{L_nabla}
\end{eqnarray}
 The four $k$--roots of the quartic dispersion relation \r{disp} are extracted as
 $k = k^{(i)}$, $k^{(ii)}$,
  $k^{(iii)}$ and   $k^{(iv)}$, where
\begin{equation}
\left.
\begin{array}{l}
k^{(i)} =    - \omega \le \sqrt{\eps \mu} + \xi \ri   \\
k^{(ii)} =  \omega \le \sqrt{\eps \mu} - \xi    \ri  \\
k^{(iii)} = - \omega \le  \sqrt{\eps \mu} - \xi \ri  \\
k^{(iv)} =  \omega \le \sqrt{\eps \mu} + \xi \ri
\end{array}
\right\}. \l{roots}
\end{equation}

The rate of energy  flow is provided by  the time--averaged
 Poynting vector  $\#P  = \frac{1}{2} \mbox{Re} \, \lec \, \#E \times \#H^* \,\ric $, where
 the
superscript $^*$ indicates the complex conjugate.
Utilizing the source--free Maxwell curl postulates, we find that for plane waves \r{pw}
propagating in the ICM specified by
\r{cr},
\begin{equation}
 \#P =
 \frac{1}{2} \exp \le -2 k_I \hat{\#k} \. \#r \, \ri \mbox{Re} \, \lec \frac{1}{\mu^*} \les  \#E_0 \times
\le
\frac{k^*}{ \omega} \hat{\#k} \times \#E^*_0  - i \xi^* \#E^*_0 \ri
 \ris \,\ric.
\end{equation}

Our particular interest in this communication
 lies in the relative orientations of the phase velocity and the power flow,
as provided by the
relative orientations of $k_R \hat{\#k} $ and  $\#P$. To
pursue this matter, let us express the components of the electric
field phasor as $\#E_0 = \le E_{0x}, E_{0y}, E_{0z} \ri$ and~---~
without loss of generality~---~choose
 $\hat{\#k} = \hat{\#z}$.
Thereby, it follows straightforwardly that
\begin{equation}
 k_R \, \hat{\#k} \. \#P = \frac{k_R}{2}
| E_{0y} |^2  \exp \le -2 k_I \hat{\#k} \. \#r \, \ri \, \mbox{Re} \, \lec \, \frac{1}{ \mu^*} \les
\frac{k^*}{\omega} \le
 1 + | \alpha |^2 \ri - i \xi^* \le \alpha  - \alpha^* \ri \ris \ric  , \l{kp}
\end{equation}
where the complex--valued quotient
\begin{equation}
\alpha = \frac{E_{0x}}{E_{0y}}\,
\end{equation}
has been introduced.
The  vector Helmholtz equation \r{Helm} yields
\begin{eqnarray}
\alpha &=& -  \frac{ \les \, \=L(i  k \, \hat{\#k}) \, \ris_{12}}{
 \les \, \=L(i  k \, \hat{\#k}) \, \ris_{11}} \,;
\end{eqnarray}
thus, the following evaluations of $\alpha$ are delivered for the
wavenumbers \r{roots}
\begin{eqnarray}
\alpha &=&  \left\{
\begin{array}{l}
i \qquad \mbox{for} \qquad k = k^{(i),(ii)} \\
- i \qquad \mbox{for} \qquad k = k^{(iii),(iv)}
\end{array} \l{alpha}
\right. .
\end{eqnarray}

Upon
combining \r{kp} with \r{alpha} we find that
\begin{equation}
 k_R \, \hat{\#k} \. \#P =
\left\{
\begin{array}{l}
\omega | E_{0y} |^2 \exp \le -2 k_I \hat{\#k} \. \#r \, \ri \,
\mbox{Re} \lec \sqrt{\eps \mu } + \xi \ric \times \mbox{Re} \lec
\sqrt{\frac{ \displaystyle \eps^*}{ \displaystyle \mu^*}} \,  \ric
  \qquad
\mbox{for} \qquad k = k^{(i),(iv)} \\
\\
\omega | E_{0y} |^2  \exp \le -2 k_I \hat{\#k} \. \#r \, \ri \,
\mbox{Re} \lec \sqrt{\eps \mu } - \xi \ric \times \mbox{Re} \lec
\sqrt{\frac{ \displaystyle \eps^*}{ \displaystyle \mu^*}} \,  \ric
 \qquad \mbox{for} \qquad k = k^{(ii),(iii)}
\end{array}
\right. .
\end{equation}
Therefore, NPV  propagation occurs provided that the following
inequalities are satisfied
\begin{equation}
\left.
\begin{array}{l}
\mbox{Re} \lec \sqrt{\eps \mu } + \xi \ric \times \mbox{Re} \lec
\sqrt{\frac{ \displaystyle \eps^*}{ \displaystyle \mu^*}} \,  \ric
< 0
  \qquad
\mbox{for} \qquad k = k^{(i),(iv)} \\
\\
\mbox{Re} \lec \sqrt{\eps \mu } - \xi \ric \times \mbox{Re} \lec
\sqrt{\frac{ \displaystyle \eps^*}{ \displaystyle \mu^*}} \,  \ric
< 0
  \qquad
\mbox{for} \qquad k = k^{(ii),(iii)}
\end{array}
\right\}. \l{NPV_cond}
\end{equation}
Finally, for a lossless ICM with $k \in \mathbb{R}$, we observe
that the NPV conditions \r{NPV_cond} reduce to
\begin{equation}
\left.
\begin{array}{l}
\sqrt{\eps \mu } < - \xi  \qquad \mbox{for} \qquad k = k^{(i),(iv)} \\
\\
\sqrt{\eps \mu } <  \xi  \qquad \mbox{for} \qquad k = k^{(ii),(iii)}
\end{array}
\right\} .
\end{equation}

\noindent{\bf 3. CONCLUDING REMARKS}

Through the introduction of magnetoelectric coupling, and the
associated expansion of the constitutive parameter space, the
scope for phase velocity and power flow to be oppositely directed
 is extended for isotropic mediums. In particular, it
is demonstrated that if the magnitude of the magnetoelectric
constitutive parameter $\xi$ is sufficiently large relative to the
permittivity $\eps$ and permeability $\mu$ then NPV propagation is
facilitated. While the parameter $\xi$ is typically small for
naturally--occurring ICMs \c{Bohren}, the NPV potential  for
artificial ICMs, designed to provide enhanced magnetoelectric
coupling, is highlighted by the results presented herein.


\begin{thebibliography}{99}
\nonumber

\bibitem{Veselago68}
V.G. Veselago, The electrodynamics of substances with
simultaneously negative values of $\eps$ and $\mu$, Sov Phys Usp
10  (1968) 509--514.

\bibitem{SSS}
R.A. Shelby, D.R. Smith and S. Schultz, Experimental verification
of a negative index of refraction, Science {292} (2001)  77--79.

\bibitem{NPV_expt1}
A. Grbic  and G.V. Eleftheriades,  Experimental verification of
backward--wave radiation from a negative index metamaterial, J
Appl Phys 92 (2002), 5930--5935.

\bibitem{NPV_expt2}
C.G. Parazzoli, R.B. Greegor, K. Li, B.E.C. Koltenbah and M.
Tanielian,
 Experimental verification and simulation of negative index of refraction
 using Snell's law,
Phys Rev Lett  90 (2003)  107401.

\bibitem{NPV_expt3}
A.A. Houck, J.B. Brock  and I.L. Chuang, Experimental observations
of a left--handed material that obeys Snell's law, Phys Rev Lett
90 (2003), 137401.

\bibitem{LM_MOTL}
A. Lakhtakia and  T.G. Mackay,
 Infinite phase velocity as the boundary between
 positive and negative phase velocities, Microwave
  Opt Technol Lett 20 (2004), 165--166.

\bibitem{LMW2}
A. Lakhtakia, M.W. McCall and W.S. Weiglhofer, Negative
phase--velocity mediums, In: W.S. Weiglhofer and A. Lakhtakia
(eds), Introduction to complex mediums for electromagnetics and
optics, SPIE Press, Bellingham, WA, USA, 2003.


\bibitem{HC02}
L. Hu  and S.T. Chui, Characteristics of electromagnetic wave
propagation in uniaxially anisotropic left--handed materials, Phys
Rev B 66 (2002), 085108.

\bibitem{Kark}
M.K. K\"{a}rkk\"{a}inen, Numerical study of wave propagation in
uniaxially anisotropic Lorentzian backward--wave slabs, Phys Rev E
68 (2003), 026602.

\bibitem{ML_PRE}
T.G. Mackay  and  A. Lakhtakia, Plane waves with negative phase
velocity in Faraday chiral mediums, Phys Rev E  69 (2004), 026602.



\bibitem{Beltrami}
A. Lakhtakia,  Beltrami Fields in Chiral Media, World Scientific,
Singapore, 1994.



\bibitem{Bohren}
C.F. Bohren, Isotropic chiral materials, In: W.S. Weiglhofer and
A. Lakhtakia (eds), Introduction to complex mediums for
electromagnetics and optics, SPIE Press, Bellingham, WA, USA,
2003.
\end{thebibliography}
\end{document}